\documentclass[letterpaper]{article} 
\usepackage{aaai25}  
\usepackage{times}  
\usepackage{helvet}  
\usepackage{courier}  
\usepackage[hyphens]{url}  
\usepackage{graphicx} 
\usepackage{natbib}  
\usepackage{caption} 
\usepackage{algorithm}
\usepackage{algorithmic}
\usepackage[frozencache,cachedir=minted-cache]{minted} 
\usepackage{tcolorbox}
\usepackage{booktabs}
\usepackage{xcolor}
\usepackage{array}
\usepackage{colortbl}
\usepackage{amsmath}
\usepackage{newfloat}
\usepackage{listings}
\DeclareCaptionStyle{ruled}{labelfont=normalfont,labelsep=colon,strut=off} 
\floatstyle{ruled}
\newfloat{listing}{tb}{lst}{}
\floatname{listing}{Listing}
\title{Uncovering LLM-Generated Code: A Zero-Shot Synthetic Code Detector via Code Rewriting}
\author{
    Tong Ye\textsuperscript{\rm 1}\equalcontrib,
    Yangkai Du\textsuperscript{\rm 1}\equalcontrib,
    Tengfei Ma\textsuperscript{\rm 2},
    Lingfei Wu\textsuperscript{\rm 3},\\
    Xuhong Zhang\textsuperscript{\rm 1}\thanks{Corresponding Author.},
    Shouling Ji\textsuperscript{\rm 1},
    Wenhai Wang\textsuperscript{\rm 1}
}
\affiliations{
    \textsuperscript{\rm 1}Zhejiang University,
    \textsuperscript{\rm 2}Stony Brook University,
    \textsuperscript{\rm 3}Anytime.AI

    {tongye,yangkaidu,zhangxuhong,sji,zdzzlab}@zju.edu.cn, tengfei.ma@stonybrook.edu, lwu@anytime-ai.com
}

\usepackage{bibentry}

\begin{document}

\maketitle

\begin{abstract}
Large Language Models (LLMs) have demonstrated remarkable proficiency in generating code. However, the misuse of LLM-generated (synthetic) code has raised concerns in both educational and industrial contexts, underscoring the urgent need for synthetic code detectors. Existing methods for detecting synthetic content are primarily designed for general text and struggle with code due to the unique grammatical structure of programming languages and the presence of numerous ``low-entropy'' tokens. Building on this, our work proposes a novel zero-shot synthetic code detector based on the similarity between the original code and its LLM-rewritten variants. Our method is based on the observation that differences between LLM-rewritten and original code tend to be smaller when the original code is synthetic. We utilize self-supervised contrastive learning to train a code similarity model and evaluate our approach on two synthetic code detection benchmarks. Our results demonstrate a significant improvement over existing SOTA synthetic content detectors, with AUROC scores increasing by 20.5\% on the APPS benchmark and 29.1\% on the MBPP benchmark.
\end{abstract}

%

\section{Introduction}
LLMs and Code LLMs have shown remarkable capability in understanding and generating code \cite{chen2021evaluating, fried2022incoder, nijkamp2022codegen,codellama2023, deepseek-coder}. Those LLMs can function as professional coding assistants for programmers, offering intelligent code completion and document generation capabilities, such as Github Copilot \cite{copilot}.

The breakthrough of LLMs has greatly improved coding efficiency and lowered the barrier to programming \cite{kazemitabaar2023novices}. However, this also raises concerns about misuse. In education, students are using LLMs to complete coding assignments and exams, making it harder to assess their true abilities \cite{kazemitabaar2023novices, denny2024computing}. A study shows GPT-4 can solve LeetCode problems at an average human level \cite{bubeck2023sparks}, increasing the risk of cheating. Moreover, LLM-generated code often contains security vulnerabilities, posing risks in industrial applications \cite{he2023large}. An evaluation found 40\% of Copilot-generated programs contain dangerous vulnerabilities \cite{pearce2022asleep}, highlighting the need for a more rigorous review of synthetic code.

\begin{figure}
    \centering
    \includegraphics[width=0.5\textwidth]{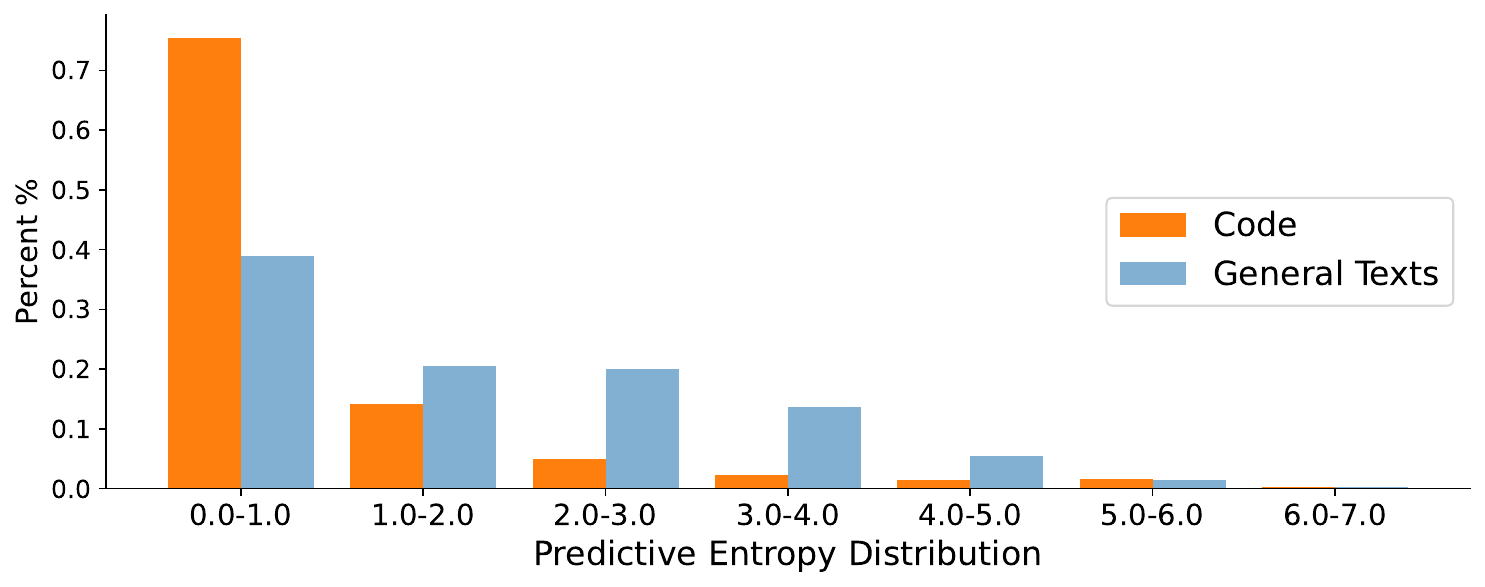}
    \caption{Token Entropy Distribution in General Texts vs. Code, Estimated by Llama-13B \cite{touvron2023llamaopenefficientfoundation}.}
    \label{fig:token-entropy}
\end{figure}

Building on these critical real-world concerns, developing a synthetic code detector is essential to address the misuse of LLM-generated code. While efforts have been made to detect LLM-generated text \cite{gehrmann-etal-2019-gltr, mitchell2023detectgpt, zellers2019defending, ippolito-etal-2020-automatic, zhong-etal-2020-neural,pu2023deepfake}, none specifically target code. The effectiveness of these methods on code content remains uncertain. Unfortunately, our experiments show that state-of-the-art detectors like GLTR \cite{gehrmann-etal-2019-gltr} and DetectGPT \cite{mitchell2023detectgpt} suffer a significant performance drop (around 32\% in AUROC) when applied to code. We thoroughly analyze the root causes of existing methods' failures, as shown below.

\begin{figure*}
    \centering
    \includegraphics[width=1.0\textwidth]{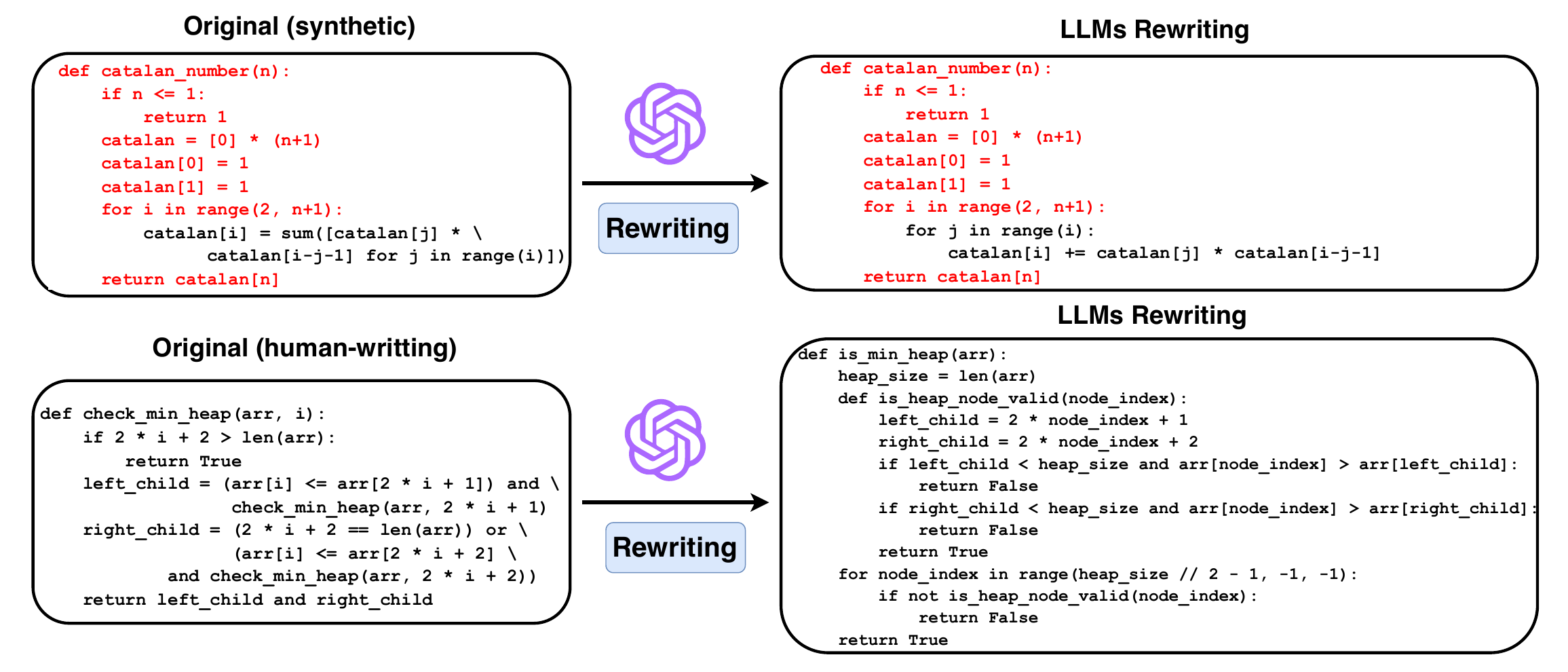}
    \caption{Rewriting Original Code (synthetic \& human-writting) with GPT-3.5-Turbo.}
    \label{fig:illustration}
\end{figure*}

\paragraph{Challenges in Applying Existing Detection Methods to Code.}

Our observations and meticulous analysis reveal that the core issue lies in the fundamental logic of existing state-of-the-art text detection methods, which rely on the statistical log probability of tokens. LLMs tend to assign higher log probabilities to the tokens they generate due to their strong confidence in these self-generated tokens. Detection methods capitalize on this trait to distinguish between model-generated and human-written text, achieving high accuracy. However, in the domain of code, this approach exposes its inherent weaknesses. Unlike natural language, code typically follows a more rigid grammatical structure. In specific programming languages and contexts, the next code token is often deterministic or has very limited options. For example, in Python, functions always start with ``def", followed by a ``:" after the function signature ``def function\_name()". These token are what we refer to as ``low-entropy" tokens. The prevalence of such ``low-entropy" tokens in code greatly reduces the effective tokens available for text detection methods to distinguish between human-written and LLM-generated code, leading to statistically insignificant differences between the two. Our investigation into the entropy distribution of code tokens versus text tokens (Figure \ref{fig:token-entropy}) confirms that the entropy distribution for text tokens is more varied compared to code, with over 70\% of code tokens having an entropy of less than 1.

\paragraph{Observations and Hypothesis.} Therefore, to effectively detect synthetic code, it is crucial to move beyond the current focus on token log probabilities and adopt a more holistic, global perspective. Based on this insight, we observed that when LLMs rewrite synthetic code—whether generated by themselves or other LLMs—they tend to produce output that closely mirrors the original code. In contrast, when rewriting human-written code, the LLM-generated output diverges more significantly. As illustrated in Figure \ref{fig:illustration}, the similarity between synthetic code and its rewritten version is striking (marked in red), while human-written code and its rewritten counterpart show more substantial differences.

This observation led to our hypothesis that the similarity between the original and rewritten code can serve as an indicator for detecting synthetic code. By focusing on code rewriting and similarity measurement, we can avoid reliance on statistical token likelihood estimation, thereby addressing the root cause of failures in existing text detectors for code. To validate this hypothesis, we propose a zero-shot synthetic code detector using Code Rewriting and Similarity Measurement. Our method consists of three crucial steps: First, we use an LLM with an appropriate prompt to rewrite a given piece of code, producing a pair of the original and rewritten code. Next, to accurately gauge the similarity between the original and rewritten code, we then develop a code similarity model that predicts a similarity score for the given code pair. Finally, using the cosine similarity metric, we estimate the expected similarity score by sampling $m$ rewritten codes following the identical procedure and averaging their scores.

During our experiments, due to the lack of existing datasets for synthetic code detection, we constructed two new datasets based on APPS \cite{hendrycks2021apps} and MBPP \cite{austin2021mbpp}. To validate the universal applicability of our hypothesis and method across different code generation tools, we generated synthetic code using two open-source LLMs (CodeLlama and StarChat) and two proprietary APIs (GPT-3.5-Turbo and GPT-4). Our experimental results demonstrate that our proposed method significantly outperforms current state-of-the-art approaches, achieving a 20.5\% improvement in AUROC on the APPS benchmark and a 29.1\% improvement on the MBPP benchmark. Additionally, our method is highly practical and requires minimal resources and permissions, only needing the ability to perform LLMs inference or access their APIs. This contrasts with previous detection methods that rely on knowledge of token log probabilities.

Overall, the main contributions are three folds:
\begin{itemize}
    \item We identify a significant performance gap when applying synthetic content detectors designed for general text to the code domain and provide a detailed analysis to uncover the reasons for this discrepancy.
    \item We propose a novel zero-shot synthetic code detection approach that leverages code rewriting and similarity measurement. This method is applicable to both open-source Code LLMs and closed-source LLMs, such as GPT-4, which only provide API access.
    \item Extensive experiments and analysis demonstrate the effectiveness of our synthetic code detection method, showing significant improvements in both accuracy and robustness compared to existing methods.
\end{itemize}

\section{Related Works}
\subsection{LLMs for Code Generation.}
With the rapid rise in popularity of LLMs, a growing number of studies have focused on Code LLMs aimed at automating software engineering \cite{chen2021evaluating, deepseek-coder}. Code generation is a crucial capability of these models, as demonstrated by pioneering works like Codex \cite{chen2021evaluating}, AlphaCode \cite{alphacode_li_2022}, CodeGeeX \cite{zheng2023codegeex}, CodeLlama \cite{codellama2023}, and GPT-4 \cite{achiam2023gpt}. 
As LLMs continue to advance, professionals across various industries are increasingly integrating them into their daily workflows. According to a community survey by \citet{stackoverflow-survey}, 44\% of experienced developers and 55\% of beginners are already using AI coding assistants, with the majority relying on two main tools: ChatGPT (83\%) and GitHub Copilot (56\%). Consequently, the automated detection of whether code is human-written or LLM-generated has become increasingly important in fields like education and industry.

\subsection{Detection of Synthetic Text.}
\label{sec:baseline}
Detecting AI-generated (synthetic) text has been a focus of research even before the emergence of LLMs. The primary approaches to synthetic text detection fall into two categories. The first treats detection as a binary classification problem, using synthetic texts from generative models to train supervised models, often based on neural networks, such as Transformers \cite{zellers2019defending, ippolito-etal-2020-automatic, zhong-etal-2020-neural, bakhtin2019real, uchendu2020authorship}. The second approach involves designing zero-shot metrics \cite{gehrmann-etal-2019-gltr, su2023detectllm, mitchell2023detectgpt} to measure the relationship between a given text and the distribution of generative models.
These methods all rely on the premise that LLMs generate tokens with higher confidence, reflected in elevated log probabilities. By statistically analyzing log probabilities, effective differentiation can be achieved. Contemporaneous study \citet{mao2024raidar} fails to reveal the “low-entropy” nature specific to code. All of the above underscores the need for a code-specific detection method from a new perspective.

\section{Methodology}
\begin{algorithm}[tb]
\caption{Zero-shot Synthetic Code Detection}
\label{alg:algorithm}
\textbf{Input}: $x$: code snippet, $\mathcal{G}$: generative model,\\
$\mathcal{M}$: similarity model, $m$: number of rewriting, $\epsilon$: threshold.
\textbf{Output}: \texttt{true}: $x$ is generated by $\mathcal{G}$, \\
                 \texttt{false}: $x$ is not generated by $\mathcal{G}$. 
\begin{algorithmic}[1] 
\STATE // generate m rewriting of $x$.
\STATE $x'_i \sim \mathcal{G}(\cdot\mid x),\,i\in [1...m]$

\STATE // estimate the similarity score.
\STATE $score \gets \frac{1}{m}\sum_i \frac{\mathcal{M}(x) \cdot \mathcal{M}(x'_i)^T}{\Vert \mathcal{M}(x) \Vert_2 \times \Vert \mathcal{M}(x'_i) \Vert_2} $
.

\IF {$score > \epsilon$}
\STATE return \texttt{true}  \quad  // $x$ is generated by $\mathcal{G}$.
\ELSE
\STATE return \texttt{false} \quad  // $x$ is not generated by $\mathcal{G}$.
\ENDIF
\end{algorithmic}
\end{algorithm}

\paragraph{Problem Definition.}
We focus on the zero-shot synthetic code detection problem, where the task is to determine whether a code snippet $x$ is generated by an LLM or written by a human. In this context, ``zero-shot" implies that we do not require a labeled dataset of synthetic and human-written code for training. Unlike previous zero-shot approaches that assume a ``white-box" setting with access to the generative model's log probability scores, we argue that this assumption is too strong for practical synthetic code detection. Many commercial LLMs provide only the generated content, without revealing the underlying log probability scores. Therefore, we investigate the more stricter ``black-box" setting, where only the generated content is available.

\paragraph{Intuition and Hypothesis.}
Distinctively, we adopt a holistic and global perspective, grounded in the intuition that every programmer develops a unique coding style, shaped by their routines and habits. Similarly, generative models, like experienced programmers, exhibit distinctive coding patterns influenced by the biases present in their training data. Therefore, LLMs can be viewed as programmers with consistent coding styles shaped by these inherent biases.

Therefore, based on consistent code writing patterns, we hypothesize that when generative LLMs are tasked with rewriting synthetic code, the differences between the rewritten and original code will be smaller compared to human-written code. Figure \ref{fig:illustration} visually illustrates this concept. Building on this hypothesis, we propose a zero-shot synthetic code detection method that leverages Code Rewriting and Similarity Measurement. The design principle is straightforward: for a given code snippet $x$, we perform multiple code rewriting using a generative LLM $\mathcal{G}$. Then, we measure the average similarity between the rewritten and original code using a code similarity model $\mathcal{M}$, trained with self-supervised contrastive learning \cite{gao2021simcse} on unlabeled code. This average similarity score serves as the detection metric. Our approach is outlined in Algorithm \ref{alg:algorithm}, with detailed explanations of Code Rewriting and Similarity Measurement in the following sections.

\begin{figure}[ht]
    \centering
    \begin{tcolorbox}[colframe=black, colback=white, width=0.49\textwidth]
        \begin{minted}[breaklines, fontsize=\footnotesize, escapeinside=||]{cpp}
### Code:
{src_code}

### Instruction:
Please explain the functionality of the given code, then rewrite it in a single markdown code block. No additional clarifications.
\end{minted}
    \end{tcolorbox}
    \caption{Prompt for Code Rewriting. }
    \label{fig:chain_of_thought_prompt}
\end{figure}

\paragraph{Code Rewriting.}
Given a code snippet $x$, we generate a rewriting of $x$ utilizing the chain of thought prompting method \cite{wei2022chainofthought}.
We prompt the LLM $\mathcal{G}$ with the original $x$, instructing it to first generate an explanation of the code, followed by a potential rewrite $x'$, as shown in Figure \ref{fig:chain_of_thought_prompt}. The intermediate code explanation can help LLM understand the original code and generate a valid rewrite according to the explanation. To minimize noise and prevent evasion tactics, we normalize $x$ by removing comments and empty lines before prompting. The LLM is instructed to return the rewritten code in Markdown format, and we remove any in-line comments during post-processing.

\paragraph{Similarity Measurement.}
\label{sec:sim}
To accurately measure the similarity between the original code and the rewritten code, we require a code similarity model $\mathcal{M}$ that can predict a similarity score $S$ for a given code pair.
Code similarity learning is a pivotal research area in AI for software engineering, often centered on code representation learning. This involves creating dense semantic representations using various code structures \cite{feng-etal-2020-codebert, ye2021misim, wang2021syncobert, guo2021graphcodebert, wang-etal-2022-code, li-etal-2022-soft} and measuring similarity through vector distances. We selected GraphCodeBERT \cite{guo2021graphcodebert} as our base model due to its widespread adoption, availability, and extensive pre-training on large-scale code data.

To enhance the code function-level representation of GraphCodeBERT, we employ self-supervised contrastive learning. Following SimCSE \cite{gao2021simcse}, we adapt the unsupervised SimCSE method to the code domain, using standard dropout as the data augmentation method for contrastive learning. For a given code snippet $x$, we take the last-layer hidden states of [CLS] token as code representation. During the unsupervised training stage, we introduce an MLP layer to obtain the final representation $\mathbf{h}$ of $x$.

Formally, for a batch of input code snippets $\{x_i\}_{i=1}^{N}$, we pass the batch through GraphCodeBERT twice, obtaining two sets of embeddings, $\{\mathbf{h}_i\}_{i=1}^{N}$ and $\{\mathbf{h}'_i\}_{i=1}^{N}$, each with a different dropout mask applied. The training objective of SimCSE is defined as:
\begin{equation}
    \mathcal{L} = - \frac{1}{N}\sum _{i=1}^{N}log\frac{e^{sim(\mathbf{h}_i,\mathbf{h}'_i) / \tau}}{\sum_{j=1}^{N} e^{sim(\mathbf{h}_i,\mathbf{h}'_j) / \tau}}
\end{equation}
Here, $\tau$ is a temperature hyperparameter set to $0.1$. The term $sim(\mathbf{h}_i,\mathbf{h}'_i)$ is the cosine similarity $\frac{\mathbf{h}_i \cdot \mathbf{h}'_i}{|\mathbf{h}_i||\mathbf{h}'_i|}$.

After the unsupervised contrastive learning stage, the MLP layer is removed, and we use only the last-layer hidden states of the [CLS] token as the code snippet representation, following \cite{gao2021simcse}.  For a given code snippet $x$ and its rewriting $x'$ generated by $\mathcal{G}$,
we obtain the final representations of $x$ and $x'$ by feeding them into the similarity model $\mathcal{M}$. It is important to note that our framework is not tied to a specific similarity model; $\mathcal{M}$ can be any implementation that effectively models code similarity, such as OpenAI's text embedding services \cite{openaiemb2022}.

Finally, we employ the cosine similarity function and estimate the expectation of the similarity score by sampling $m$ rewritten code, formally expressed as: 
\begin{equation}
    score(x) = \mathrm{E}_{x' \sim \mathcal{G}(\cdot|x)} sim(\mathcal{M}(x),\mathcal{M}(x'))
\end{equation}
While a larger $m$ can provide a more accurate estimation, it also requires generating more rewrites. Excitingly, our experiments show that using just 4 rewrites is sufficient to achieve excellent detection performance.

\section{Experiment Setting}
\subsection{Benchmarks.} Considering the absence of existing benchmarks for evaluating synthetic code detectors, we build two synthetic code detection benchmarks in Python using APPS \cite{hendrycks2021apps} and MBPP \cite{austin2021mbpp}. The choice of Python as the programming language for constructing our benchmark stems from several key considerations. Firstly, a significant portion of training data for prevalent code generation models like CodeGen \cite{nijkamp2022codegen}, Incoder \cite{fried2022incoder}, and StarCoder \cite{li2023starcoder} is constituted by Python code. These models exhibit notably superior code generation capabilities in Python than in other languages. Consequently, the challenge and significance of detecting code generated by LLMs are most pronounced within the context of Python. Secondly, popular datasets used for code generation evaluation, such as HumanEval \cite{chen2021evaluating}, MBPP \cite{austin2021mbpp}, and APPS \cite{hendrycks2021apps}, primarily provide Python test cases. We adopted Python as our chosen language to align with this prevailing practice. Furthermore, Python ranks among the most widely used programming languages \cite{stackoverflow-developer}, adding practical significance to validating our detection performance. Specifically, for the generation of synthetic code, we utilize the natural language description of each sample as a prompt and employ four widely-used code generation tools: CodeLlama \cite{codellama2023}, StarChat \cite{Tunstall2023starchat-alpha}, GPT-3.5-Turbo \cite{openai2022gpt35} and GPT-4 \cite{achiam2023gpt}. 

\paragraph{Choice of Generation Tools.}
We select two open-sourced Code LLMs, CodeLlama-13B-Instruct \cite{codellama2023} and StarChat-Alpha \cite{Tunstall2023starchat-alpha}, along with two proprietary generation APIs: GPT-3.5-Turbo \cite{openai2022gpt35} and GPT-4 \cite{achiam2023gpt}. According to the survey \cite{stackoverflow-survey} conducted by StackOverflow, GPT-3.5-Turbo is the most extensively used tool in daily programming. GPT-4 represents the state-of-the-art in code generation tools to date. Consequently, we have chosen these two models to typify proprietary generation APIs. CodeLlama and StarChat are two potent and popular open-sourced code generation tools with millions of downloads on HuggingFace. We opted for the popular 13B and 15.5B versions due to resource constraints. 

\paragraph{Constructing APPS benchmark.}
APPS \cite{hendrycks2021apps} is a benchmark for code generation. Each sample in APPS 
contains problem descriptions and submitted solutions.
The original APPS comprises 5,000 test data and 5,000 training data instances. In our construction, we exclusively utilized the test data. The rationale for excluding the training data stems from concerns that it might have been incorporated into the training datasets of LLMs. Additionally, the APPS test data is widely used by academia to assess the code generation capabilities of LLMs \cite{codellama2023,li2023starcoder,deepseek-coder} and is typically excluded from their training dataset. In our detection task, even if the APPS test data were included in the training datasets of LLMs, it wouldn't pose significant issues. This situation would cause the generated code to closely resemble human-written code, making detection task more challenging. Nevertheless, our method still performs well.

Within the pool of 5,000 test data instances, we removed those with web hyperlinks in the Markdown format, as these links could affect code quality and are rarely used by developers. This reduced the dataset to 3,846 instances. We then randomly selected 1,540 instances (40\%) from this refined set and used 770 of them for synthetic code generation due to resource constraints. For those 770 problems, we generate synthetic code using chain of thought prompting \cite{wei2022chainofthought} with four generation tools. The exact prompt used is shown in Figure \ref{fig:apps-gen-prompt}, with a generation temperature set to $0.7$ and $top\_p$ set to 0.95. 
For the remaining 770 problems (1540 minus 770), we randomly sample a solution from all valid solutions as human-written code. The detailed dataset statistics are listed in Table \ref{tb:data-statistics-apps}. Due to economic costs, we only had GPT-4 generate 100 instances.

\begin{table}[!ht]
    \centering
    \small
    \begin{tabular}{lcccc}
    \toprule
    Generator  & CodeLlama & StarChat & GPT-3.5 & GPT-4 \\
    \midrule
    \# Sample & 770/770 & 770/770 & 770/770 & 100/100 \\
    \# Char & 630/458 & 630/456 & 630/420 & 440/401 \\
    \# Line & 19.1/15.0 & 19.1/14.6 & 19.1/14.7 & 18.7/13.6 \\
     \toprule
    \end{tabular}
\caption{APPS Benchmark Statistics. We count the averaged number of chars and lines for human-written/synthetic code.}
\label{tb:data-statistics-apps}
\end{table}

\paragraph{Constructing MBPP benchmark.}
MBPP \cite{austin2021mbpp} is another benchmark for evaluating AI code generation. It contains 1,000 crowd-sourced Python programming problems designed to be solvable by entry-level programmers, covering programming fundamentals, standard library functionality, etc. Compared to APPS, MBPP's problems are more straightforward and frequently encountered in daily programming practice. We follow the same construction process as APPS. The detailed dataset statistics are listed in Table \ref{tb:data-statistics-mbpp}.

\begin{table}[!ht]
    \centering
    \small
    \vspace{-3mm}
\begin{tabular}{lcccc}
\toprule
Generator  & CodeLlama & StarChat & GPT-3.5 & GPT-4 \\
\midrule
\# Sample & 233/233 & 233/233  & 233/233 & 100/100 \\
\# Char & 256/192 & 256/251  & 256/265 & 267/238 \\
\# Line & 10.0/6.6 & 10.0/8.2  & 10.0/9.2 & 10.4/8.4 \\
 \toprule
\end{tabular}
\caption{MBPP Benchmark Statistics.}
\label{tb:data-statistics-mbpp}
\end{table}

\subsection{Metrics.} We evaluate all detectors on our benchmarks using the Area Under the Receiver Operating Characteristic (AUROC) curve, following \citet{mitchell2023detectgpt}.

\subsection{Baselines.}
To demonstrate the effectiveness of our zero-shot synthetic code detector, we adapt the following zero-shot detection methods utilizing a surrogate score model to approximate the true distribution of the generative model: \textbf{log P(x)}  \cite{gehrmann-etal-2019-gltr}, \textbf{LogRank}, \textbf{Rank}, \textbf{Entropy} \cite{mitchell2023detectgpt}, \textbf{LRR}, \textbf{NPR} \cite{su2023detectllm} and \textbf{DetectGPT} \cite{mitchell2023detectgpt}. To adapt these detection methods to code content and effectively approximate the true distribution of the generative Code LLM, the surrogate score model should be replaced with LLMs trained on large-scale code content. In this study, we examine two open-sourced Code LLMs: StarChat-Alpha and CodeLlama.
We also select two supervised detectors and compare them to the zero-shot detectors: \textbf{GPTZero} \cite{gptzero}, a leading AI content detection services, is trained on millions of synthetic texts sampled from various generative models, including ChatGPT, GPT-4 and Bard. 
\textbf{OpenAI-Detector} \cite{openai_detector}, an open-sourced detector, is trained on texts sampled from GPT-2 and is based on Roberta-large \cite{liu2019roberta}.

\paragraph{\textbf{log P(x)}}
This method calculates the average log probability of each token. Code snippets with higher log\ P(x) scores are more likely to be synthetic \cite{gehrmann-etal-2019-gltr}.

\paragraph{\textbf{LogRank} and \textbf{Rank}} 
These metrics use the averaged (log-) rank of tokens in the predicted distribution as a detection metric. Code snippets with lower ranks tend to be sampled from AI models.

\paragraph{\textbf{Entropy}} 
We use the averaged predictive entropy of each token as another baseline following \citet{mitchell2023detectgpt}. Synthetic code will be more ``in-distribution" for the generative model, leading to more confident predictions.

\paragraph{\textbf{LRR} and \textbf{NPR}} 
These two metrics, introduced by \citet{su2023detectllm}, are used in our study with the identical calculation formula as in the original paper without any modifications, ensuring fairness.
\begin{equation}
\text{LRR} = \left| \frac{\frac{1}{t}\sum_{i=1}^t\log p_{\theta}(x_i|x_{<i})}{\frac{1}{t}\sum_{i=1}^t\log r_{\theta}(x_i|x_{<i})} \right| 
\end{equation}
where $r_{\theta}(x_i|x_{<i})\geq 1$ is the rank of token $x_i$ conditioned on the previous tokens.
\begin{equation}
    \text{NPR} = \frac{\frac{1}{n}\sum_{p=1}^n\log r_{\theta}(\tilde{x}_p)}{\log r_{\theta}(x)}
\end{equation}
where small perturbations are applied on the target text $x$ to produce the perturbed text $\tilde{x}_p$.

\paragraph{\textbf{DetectGPT}} 
DetectGPT \cite{mitchell2023detectgpt} uses the average token probability disparity between the given text and perturbed texts to detect whether the given text is located at the probability curvature of the generative model. To better adapt DetectGPT for the code domain, we made minor modifications. Specifically, we replace the original perturbation model, T5-large \cite{raffel-t5} with CodeT5-large \cite{wang2021codet5}. All other settings are the same as in the original implementation.

\begin{table*}[!ht]
    \centering
    \small
\begin{tabular}{lccccc|ccccc}
\toprule
Dataset   & \multicolumn{5}{c}{\textbf{APPS}}                     & \multicolumn{5}{c}{\textbf{MBPP}}                     \\ \cmidrule(lr){2-6} \cmidrule(lr){7-11}
Generators   & CodeLlama & StarChat & GPT-3.5 & GPT-4 & \textbf{Avg.} & CodeLlama & StarChat & GPT-3.5 & GPT-4 & \textbf{Avg.} \\ \midrule
GPTZero   & 52.71     & 56.25    & 53.68   & 58.24 & 55.22   & 59.53     & 60.82    & 57.29   & 61.55 & 59.80    \\
OpenAI   & 56.32     & 50.08    & 48.48   & 55.81 & 52.67   & 48.81     & 47.40    & 43.31   & 46.44 & 46.49    \\ \hline
\rowcolor{gray!20} \multicolumn{11}{c}{Using CodeLlama as Detector LLM}                                    \\
log P(x)  & 66.14     & 59.40    & 64.58   & 59.27 & 62.35   & 50.70     & 53.84    & 63.05   & 53.35 & 55.24    \\

LogRank     & 69.79   & 61.54    & 67.31   & 62.89 & 65.38   & 60.76    & 58.56    & 68.05   & 58.91 & 61.57     \\
Rank        & 52.17     & 48.63    & 50.77   & 48.04 & 49.90   & 25.99     & 35.75    & 42.03   & 36.33 & 35.03     \\
Entropy     & 58.91     & 54.71    & 61.49   & 55.87 & 57.75   & 37.22     & 44.90    & 50.88   & 43.40 & 44.10    \\
DetectGPT   & 61.28     & 57.71    & 62.06   & 53.41 & 59.85   & 56.28     & 53.18    & 66.56   & 63.84      & 59.96     \\
LRR       & 67.15 & 62.16 & 67.82 & 60.06 & 64.30 & 53.25 & 56.32 & 64.29 & 54.62 & 57.12 \\
NPR & 65.49     & 60.08    & 66.53   & 58.62 & 62.68   & 54.37    & 55.10   & 68.85   & 64.96    & 60.82     \\
\cmidrule(lr){1-6} \cmidrule(lr){7-11}
Ours $m=2$  & 80.78     & 72.91    & 73.12   & 68.19 & 73.75   & 77.90     & 68.88  & 76.36   & 75.02 & 74.54     \\
Ours $m=4$  & 85.42     & 76.53    & 77.70   & 74.29 & 78.49     & 82.91     & 71.50  & 79.83   & 77.71 & 77.99     \\
    Ours $m=8$  & \textbf{87.77}     & 78.13    & 80.23   & 74.51 & 80.16   & \textbf{86.21}     & 75.70  & 83.58   & 81.75 & \textbf{81.81}   \\ \hline
\rowcolor{gray!20} \multicolumn{11}{c}{Using StarChat as Detector LLM}                                     \\
log P(x)  & 66.41     & 65.27    & 65.54   & 62.18 & 64.85   & 55.81     & 64.86    & 69.91   & 60.17 & 62.69    \\
LogRank     & 66.95     & 65.81    & 66.74   & 64.25 & 65.93   & 58.69     & 65.31    & 69.56   & 59.55 & 63.28    \\
Rank        & 53.85     & 48.24    & 50.37   & 49.77 & 50.56   & 37.24     & 44.48    & 47.97   & 46.47 & 44.04     \\
Entropy     & 56.55     & 55.43    & 59.60   & 55.30 & 56.72   & 39.03     & 48.72    & 55.35   & 47.22 & 47.33    \\
DetectGPT   & 60.92     & 58.23    & 61.52   & 58.62 & 61.26   & 54.41     & 55.74    & 66.49   & 65.02     & 60.42     \\
LRR       & 66.55 & 68.91 & 68.45 & 65.88 & 67.45 & 56.80 & 66.74 & 69.87 & 60.16 & 63.39 \\
NPR & 64.47     & 63.60    & 66.43    & 65.00 & 64.88   & 54.20     & 60.00    & 70.43 & 66.17    & 62.70     \\
\cmidrule(lr){1-6} \cmidrule(lr){7-11}
Ours $m=2$  & 81.93     & 77.23    & 72.46   & 72.89 & 76.13   & 79.68     & 73.79    & 79.28   & 69.24 & 75.50     \\
Ours $m=4$  & 85.51     & 79.24    & 74.58   & 77.35 & 79.17   & 80.61     & 76.44    & 81.05   & 74.67 & 78.19    \\
Ours $m=8$  & 87.24     & 81.35    & 76.28   & 77.84 & 80.68   & 83.67     & 79.00    & 83.17   & 78.04 & 80.97     \\ \hline

\rowcolor{gray!20}  \multicolumn{11}{c}{Using GPT-3.5-Turbo as Detector LLM} \\
Ours $m=2$  & 77.84    & 81.67     & 79.02   & 79.04 & 79.39   & 66.21     & 77.29   & 83.05   & 83.45 & 77.50    \\
Ours $m=4$  & 78.21    & 82.22     & 82.12   & 78.69 & 80.31   & 67.00     & 78.87   & 85.39   & 82.23 & 78.37     \\
Ours $m=8$  & 78.47    & \textbf{82.48}     & \textbf{83.25}   & \textbf{80.87} & \textbf{81.27}   & 67.66     & \textbf{79.23}   & \textbf{86.23}   & \textbf{84.00} & 79.28    \\
\bottomrule
\end{tabular}
\caption{Main Results. The first two rows list the two benchmarks and their corresponding four generation tools. The subsequent sections detail the AUROC scores of our methods, compared to seven other zero-shot detectors, using three Detector LLMs.
}
\label{tab:main-results}
\end{table*}

\paragraph{\textbf{GPTZero}} It is a proprietary detection API \cite{gptzero}. We request the provided API with the input code following their official documentation. We use the ``completely\_generated\_prob" field returned by their API as the detection score for AUROC calculation. ``completely\_generated\_prob" refers to the probability that the entire code snippet is generated by an AI model.

\paragraph{\textbf{OpenAI-Detector}} The finetuned Roberta-large model checkpoint provided by the authors is used for detection \cite{openai_detector}. This model is a binary classifier, and we use the ``$log\_softmax$" score of the final output layer as the detection score for AUROC calculation.

\subsection{Detector (Rewriting) LLMs.} 
Our approach necessitates a rewriting LLM, $\mathcal{G}$, to serve as the detector LLM. 
We evaluate CodeLlama, StarChat-Alpha, and GPT-3.5-Turbo as detector LLMs for our method while using CodeLlama and StarChat-Alpha for baseline comparison. For code rewriting, we utilize nucleus sampling with a top-$p$ of $0.95$ and a temperature of $0.8$.

\subsection{Similarity Model Training.} 
We continue to train a code similarity model by exploiting unsupervised SimCSE from initial GraphCodeBERT. In the SimCSE training phase, we collected 160k code snippets from CodeSearchNet \cite{husain2020codesearchnet} Python subset and 140k code snippets from submitted solutions of APPS as training data. We meticulously reviewed and ensured that there is no overlap between the 140k code snippets and the constructed APPS benchmark. We use the Adam optimizer and set the maximum learning rate to 1e-4 with linear decay. We use 4 NVIDIA 3090s with batch size 16 on each GPU and train the model for 5 epochs. In addition to the GraphCodeBERT model, we also experimented with the UniXcoder \cite{guo-etal-2022-unixcoder} as another similarity model by following the same SimCSE training process.

\section{Experiment Results}
\label{main_results}
\paragraph{Main Results.} We list the AUROC score of all zero-shot detectors on the APPS and MBPP benchmarks in Table \ref{tab:main-results}. Among the seven baselines, LogRank, LRR, and NPR are the most effective methods. However, the performance of baseline zero-shot detectors, including log P(x), DetectGPT, LRR, and NPR, drops significantly on code benchmarks compared to their reported performance on general text detection \cite{mitchell2023detectgpt, su2023detectllm}. This decline is consistent across different generation tools and detector LLMs. The primary reason, as mentioned earlier, is that SOTA zero-shot text detectors rely on token log probabilities, which are less effective for code due to their uniform grammatical structure and massive ``low entropy" tokens. Additionally, we observe that baseline performance improves when the Detector LLM and generation tool are identical, but declines when they differ.

As shown in Table \ref{tab:main-results}, our proposed methods significantly outperform previous approaches across all four code generation tools, with a 20.5\% improvement on the APPS benchmark and a 29.1\% improvement on the MBPP benchmark. This success is largely due to our holistic viewpoint to detecting synthetic code, which moves beyond token-wise scoring. By employing code rewriting and similarity measurement, our method effectively addresses the limitations of previous techniques in the code domain, leveraging the consistent coding style of LLMs and their tendency to generate the most confident, likely code. Notably, our method achieves strong detection performance with just two rewritings for estimating similarity. Increasing the number of rewritings ($m = 4$ or $m = 8$) further enhances performance.

\paragraph{Effectiveness of Detector LLM.} 
We observe that GPT-3.5-Turbo performs optimally as the detector LLM for identifying code generated by StarChat, GPT-3.5-Turbo, and GPT-4, but not for CodeLlama. The best detection results for CodeLlama-generated code are achieved when using CodeLlama itself as the detector. This is likely because StarChat is fine-tuned on an instruction dataset distilled from GPT-3.5-Turbo and GPT-4, leading to a closer distribution among these three models. In contrast, CodeLlama, trained on a dataset derived from self-instructing Llama-2 \cite{touvron2023llama}, exhibits a different distribution, explaining GPT-3.5-Turbo's suboptimal performance in detecting CodeLlama-generated code.
Regarding the superior performance of GPT-3.5-Turbo over StarChat in detecting StarChat-generated code, we found that StarChat tends to oversimplify rewrites for complex problems, whereas GPT-3.5-Turbo provides more accurate and complete rewrites.

\paragraph{Comparison to Supervised Detector.}
The first two rows in Table \ref{tab:main-results} display the results of the supervised detectors, GPTZero and OpenAI-Detector. Despite being trained on millions of labeled samples, GPTZero and OpenAI-Detector perform no better than random guessing when applied to code content. This suggests that these supervised detectors may overfit to their training distribution and struggle to generalize to new domains without adaptive tuning. In contrast, our proposed zero-shot detection methods show strong generalization across different code distributions and outperform the supervised detectors.

\paragraph{Ablation Study.}
We conducted an ablation experiment to assess the contributions of the two primary components in our design: Code Rewriting and Similarity Measurement. We considered two ablation settings: First, replacing the Code Rewriting with in-fill perturbation following DetectGPT \cite{mitchell2023detectgpt} while retaining the Similarity Measurement (Sim); Second, retaining Code Rewriting (CR) and replacing the Similarity Measurement with token-wise score difference (LogProb, LogRank, Rank and Entropy). The results are presented in Table \ref{tab:ablation-study-apps}. 
\begin{table}[!ht]
    \centering
    \small
\begin{tabular}{lcccc}
\toprule
Dataset   & \multicolumn{4}{c}{APPS} \\ 
\cmidrule(lr){2-5}
Generators   & CodeLlama & StarChat & GPT-3.5 & GPT-4 \\ 
\midrule
CR + LogProb    & 57.29          & 46.70          & 50.58          & 55.13          \\
CR + LogRank    & 75.28          & 64.76          & 75.85          & 67.69          \\
CR + Rank      & 62.34           & 55.42          & 60.59          & 59.34                \\
CR + Entropy    & 60.39          & 59.42          & 68.20          & 59.35          \\ 
Perturb + Sim   & 74.55          & 71.55          & 74.23          & 53.43  \\ 
\midrule
CR + Sim & \textbf{87.77} & \textbf{76.96} & \textbf{80.23} & \textbf{74.51} \\ 
\bottomrule
\end{tabular}
\caption{Ablation Study. All results are reported when using CodeLlama as the Detector LLM.}
\label{tab:ablation-study-apps}
\end{table}

The ablation results indicate that our ``CR + Sim" method significantly enhances the detection performance. Replacing either Code Rewriting or Similarity Measurement leads to a noticeable decrease in performance, underscoring the importance of these two components.

\section{Deeper Analysis}
To further explore the effectiveness and robustness of our methods, we conducted comprehensive experiments to evaluate performance across various settings and scenarios. All experiments in this section are conducted using GPT-3.5-Turbo as the generation tool.

\paragraph{Choice of Similarity Model.}
In the Methodology section, we exploit self-supervised contrastive learning to train a better code similarity model based on GraphCodeBERT (GCB-SimCSE). To investigate the impact of the Similarity Model, we experiment with three other variants of code similarity models: UniXcoder \cite{guo-etal-2022-unixcoder} trained by SimCSE (Unix-SimCSE), the original GraphCodeBERT model with average pooling (GCB-avg) and OpenAI's text embedding service (Text-ada-002) \cite{openaiemb2022}. The results on the APPS benchmark are listed in Table \ref{tab:ablation-apps}. While Unix-SimCSE achieves the highest performance, the other three models still outperform previous zero-shot detectors. 

\begin{table}[!ht]
    \centering
    \small
    \begin{tabular}{lccc}
         \toprule
         Detectors & CodeLlama & StarChat & GPT-3.5 \\
         \midrule
         Unix-SimCSE & \textbf{81.21} & \textbf{77.84} & \textbf{86.37}  \\
         GCB-SimCSE & 80.23 & 76.28 & 83.25  \\
         GCB-avg    & 72.79 & 70.95  & 72.22  \\
         Text-ada-002   & 70.18 & 69.29 & 76.48  \\
        \bottomrule
    \end{tabular}
    \caption{Choice of different similarity models. We use CodeLlama, StarChat and GPT-3.5-turbo as Detector LLMs.}
    \label{tab:ablation-apps}
\end{table}

These findings suggest that our method is not reliant on a single similarity model. Additionally, since UniXcoder is an enhanced version of GraphCodeBERT, the superior performance of Unix-SimCSE over our original GCB-SimCSE indicates that detection performance can be further improved by using a stronger and more robust similarity model.

\begin{figure*}[!ht]
    \centering
    \includegraphics[width=1.0\textwidth]{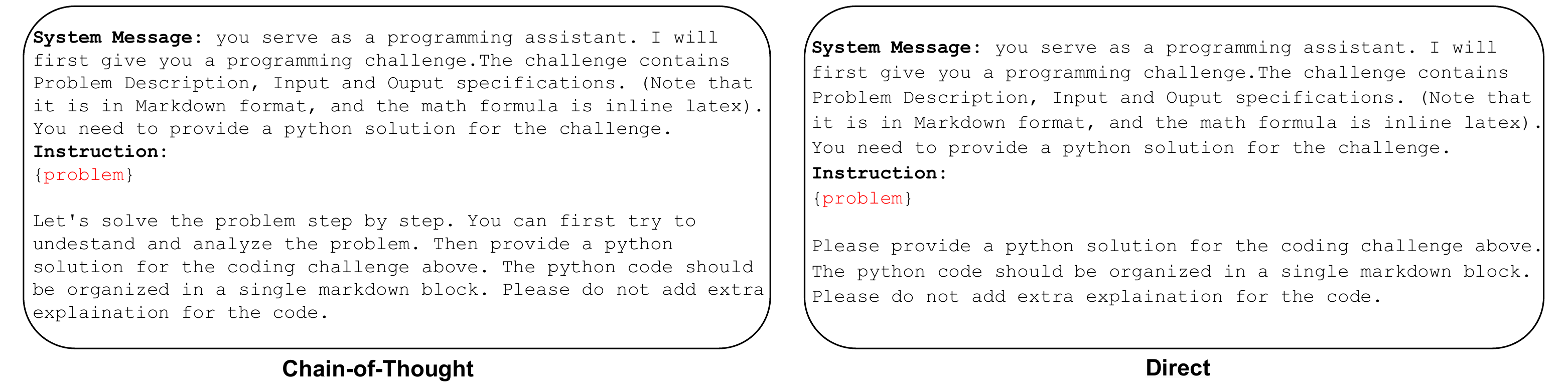}
    \caption{Prompt templates for synthetic code generation. We use chain-of-thought prompting for reporting main results. We use direct prompting as part of model analysis for the impact of generation prompts.}
    \label{fig:apps-gen-prompt}
\end{figure*}

\begin{figure}[!h]
    \centering
    \includegraphics[width=0.5\textwidth]{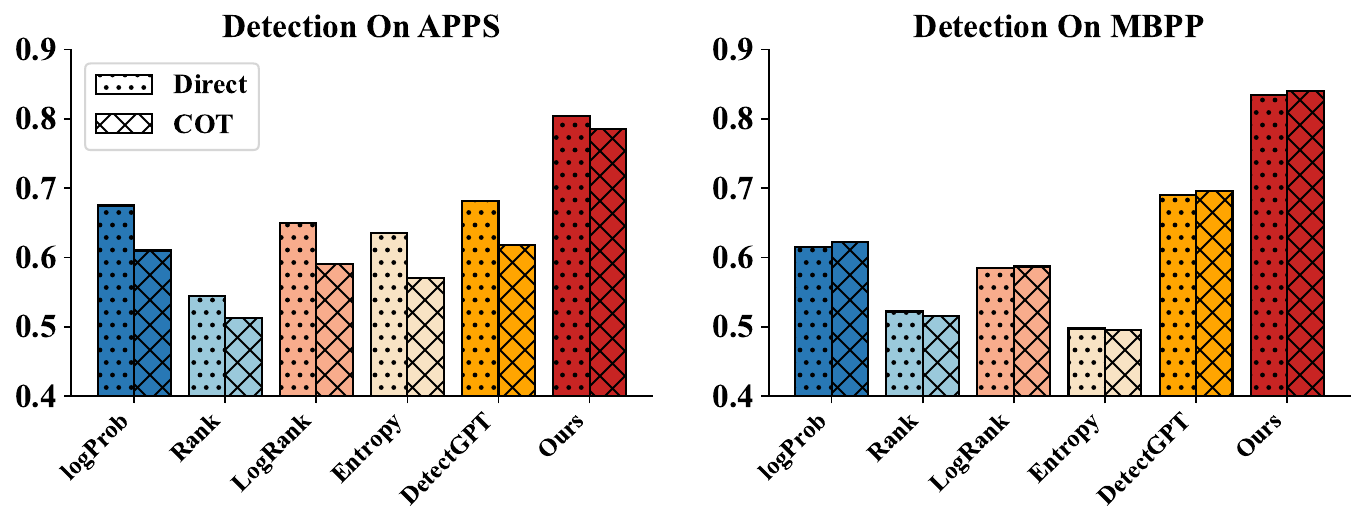}
    \caption{Impact of generation prompts.}
    \label{fig:vary_gen_prompt}
\end{figure}

\paragraph{Impact of Generation Prompts} 
It is widely acknowledged that LLM outputs can be notably influenced by prompts. Therefore, the prompts used to generate synthetic code also have a substantial impact on the generated output. To assess this impact, we modified the prompts used in our benchmark for synthetic code generation. Specifically, we compared the effectiveness of two prompting approaches: chain of thought prompts and direct prompts. The latter directly asks the model to generate a solution without preliminary analysis. 
The difference between these two prompts is illustrated in Figure \ref{fig:apps-gen-prompt} and the results (AUROC) under two different prompts are presented in Figure \ref{fig:vary_gen_prompt}.

The results in Figure \ref{fig:vary_gen_prompt} indicate that our proposed methods are more robust than previous zero-shot detectors when the generation prompts are varied across both benchmarks.

\begin{figure}[!ht]
    \centering
    \includegraphics[width=0.5\textwidth]{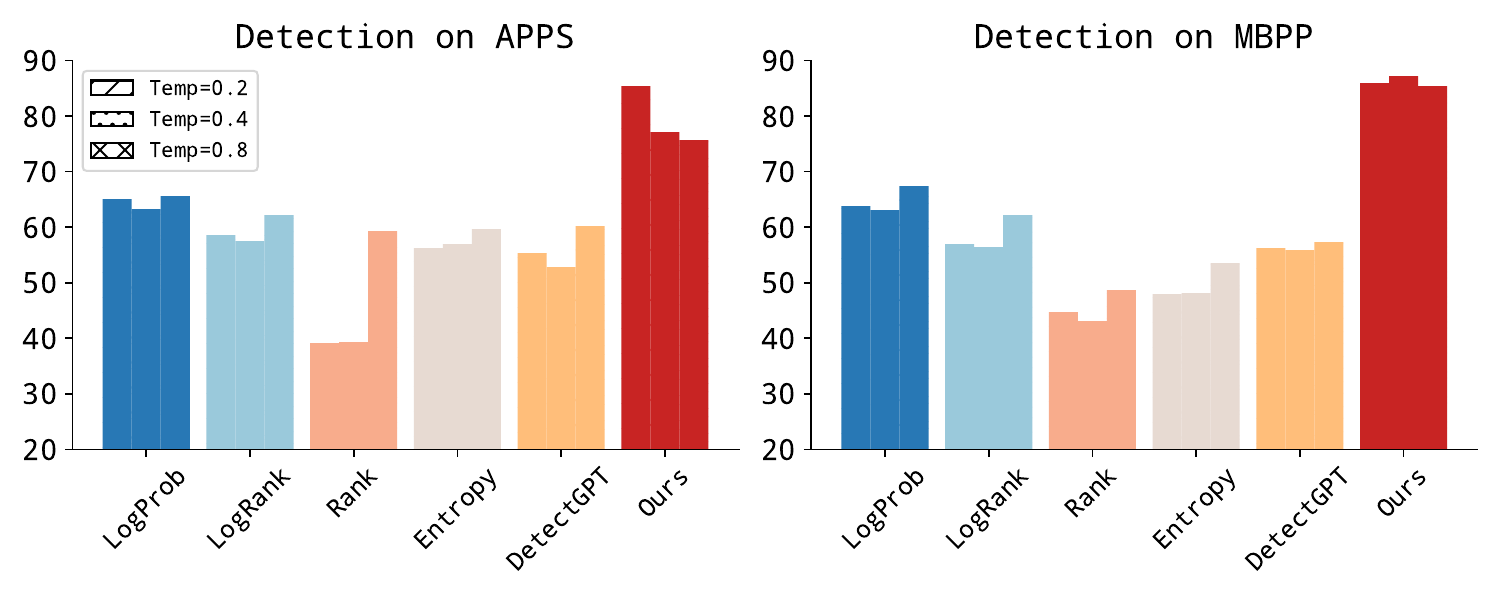}
    \caption{Impact of Decoding Strategy.}
    \label{fig:vary_decode_temp}
\end{figure}

\paragraph{Impact of Decoding Strategy.}
Adjusting the temperature parameter in LLMs balances output diversity and accuracy, with lower temperatures yielding more deterministic and consistent results, while higher temperatures produce more varied and creative outputs. To investigate this,
we conducted experiments on the APPS and MBPP benchmarks, varying the generator temperature from $[0.2, 0.4, 0.8]$ while keeping the rewriting temperature fixed at $0.8$. This range was chosen based on the Codex \cite{chen2021evaluating}, which found $T=0.2$ and $T=0.8$ optimal for pass@1 and pass@100 rates, respectively. The results in Figure \ref{fig:vary_decode_temp} demonstrate that our method exhibits superior consistency across different temperatures.

\paragraph{Generalizability to Different Programming Languages.}

To assess the generalizability of our method to different programming languages, we construct an additional C++ benchmark using the Code Contest dataset \cite{alphacode_li_2022}. We generate synthetic code using the same four-generation tools. The detection performance on C++ is presented in Table \ref{tab:cpp-results}. Our method still achieves notable improvement compared to other zero-shot baselines on C++ benchmarks. Moreover, we observe that synthetic C++ code is easier to detect for both our method and other baselines compared to Python, suggesting a more significant distribution gap between synthetic and human-written C++ code. 
\begin{table}[!ht]
    \centering
    \small
\begin{tabular}{lcccc}
\toprule
Dataset   & \multicolumn{4}{c}{Code Contest C++} \\ 
\cmidrule(lr){2-5}
Generators   & CodeLlama & StarChat & GPT-3.5 & GPT-4 \\ 
\midrule
log P(x) & 67.82  &61.92 &73.55 & 69.40 \\
LogRank & 59.35 & 57.12 & 66.68 & 62.43 \\
Rank & 54.67 & 52.19  & 60.29 & 54.21 \\
Entropy & 43.83 & 39.52 & 53.74 & 55.50 \\
DetectGPT & 62.99 & 60.45 & 75.47 & 64.05 \\ \midrule
Ours $m = 8$ & \textbf{89.87} & \textbf{83.42} & \textbf{90.82} & \textbf{88.85} \\
\bottomrule
\end{tabular}
\caption{Detection Results on C++. All results are reported when using CodeLlama as the Detector LLM.}
\label{tab:cpp-results}
\end{table}

\begin{figure}[!ht]
    \centering
    \includegraphics[width=0.50\textwidth]{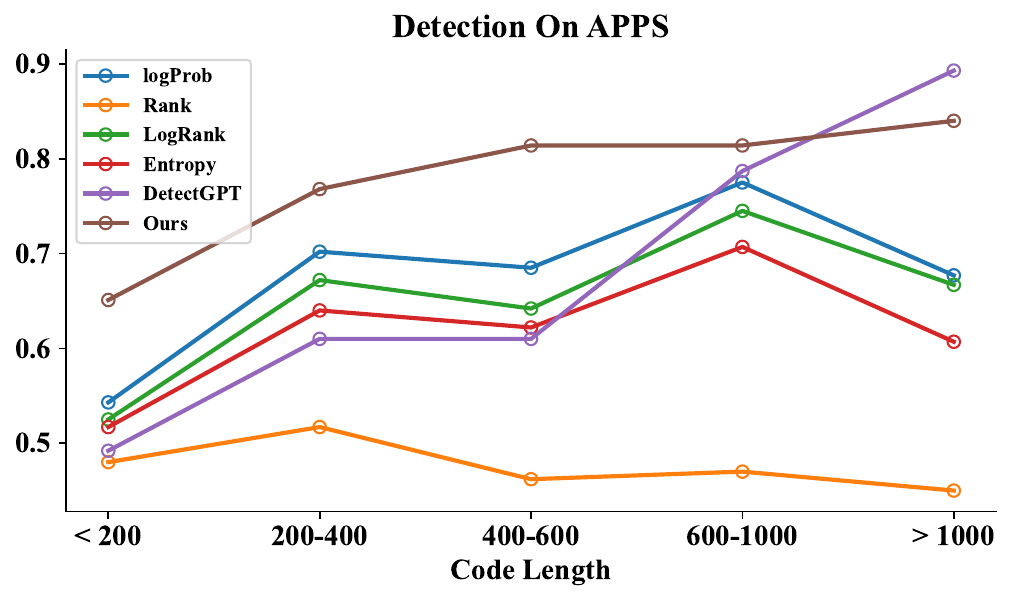}
    \caption{Impact of Code Length.}
    \label{fig:vary_codelength}
\end{figure}

\paragraph{Impact of Code Length.}
We also investigate how the length of the code snippet impacts the detection performance of the zero-shot synthetic code detectors. We divide the samples in the APPS benchmark into five groups according to the code string length. Each group has an equal number of synthetic and human-written code snippets. The results are plotted in Figure \ref{fig:vary_codelength}.

In general, the performance of all detectors improves as the code length falls within the 0-1000 character range. 
DetectGPT's performance surpasses ours when the code exceeds 1000 characters. However, DetectGPT is inferior to the other baselines for code lengths less than 600, the range where the majority ($70\%$) of code snippets lie. Our method displays strong performance on all ranges and is more robust to changes in code length when compared to other baselines.

\begin{figure}[!h]
    \centering
    \includegraphics[width=0.5\textwidth]{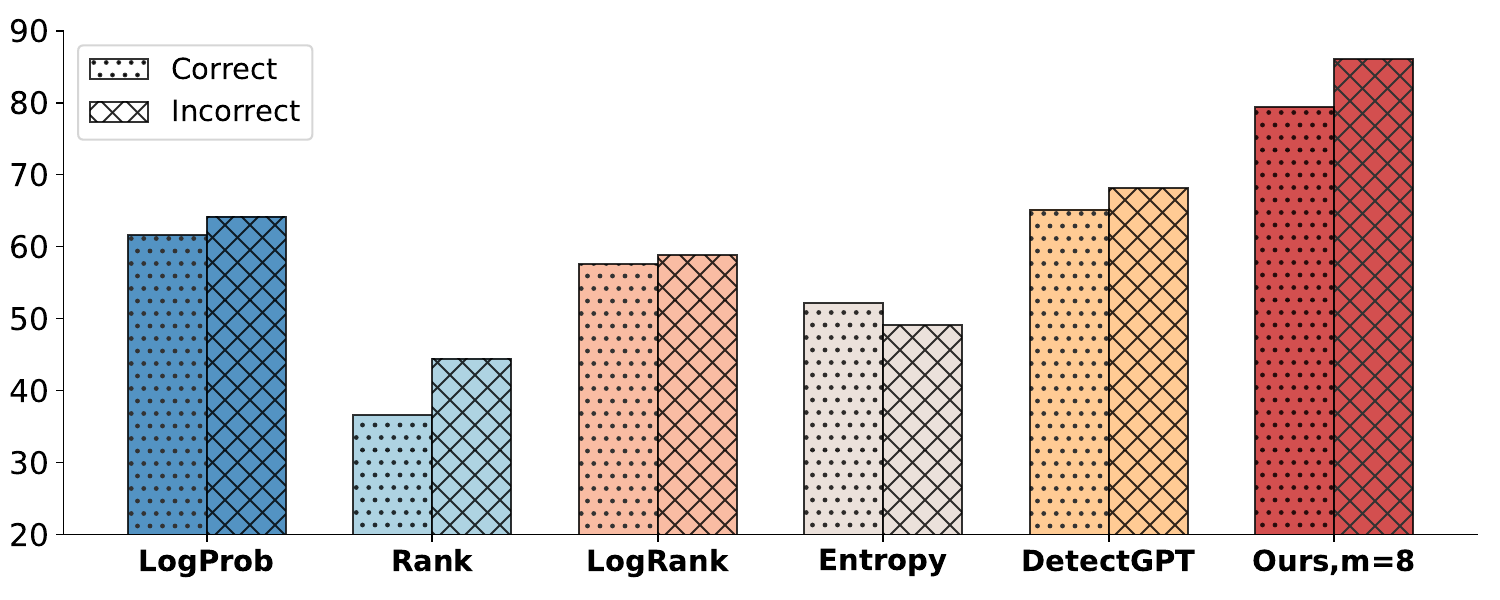}
    \caption{Impact of code correctness.}
    \label{fig:code_correctness}
\end{figure}

\paragraph{Impact of Code Correctness.}
We consider that the distribution of correct code solutions likely differs from incorrect ones since the correct synthetic code may much closer to human-written code, making it more difficult to detect. To explore the impact of code correctness, we separately present the detection AUROC for correct and incorrect codes on the MBPP benchmark using CodeLlama as Detector LLM, as shown in Figure \ref{fig:code_correctness}. The results indicate that detecting correct code is more challenging than detecting incorrect code, both with our method and baselines. Nonetheless, our method consistently outperforms baselines when dealing with correct code.

\begin{figure}[!ht]
    \centering
    \includegraphics[width=0.5\textwidth]{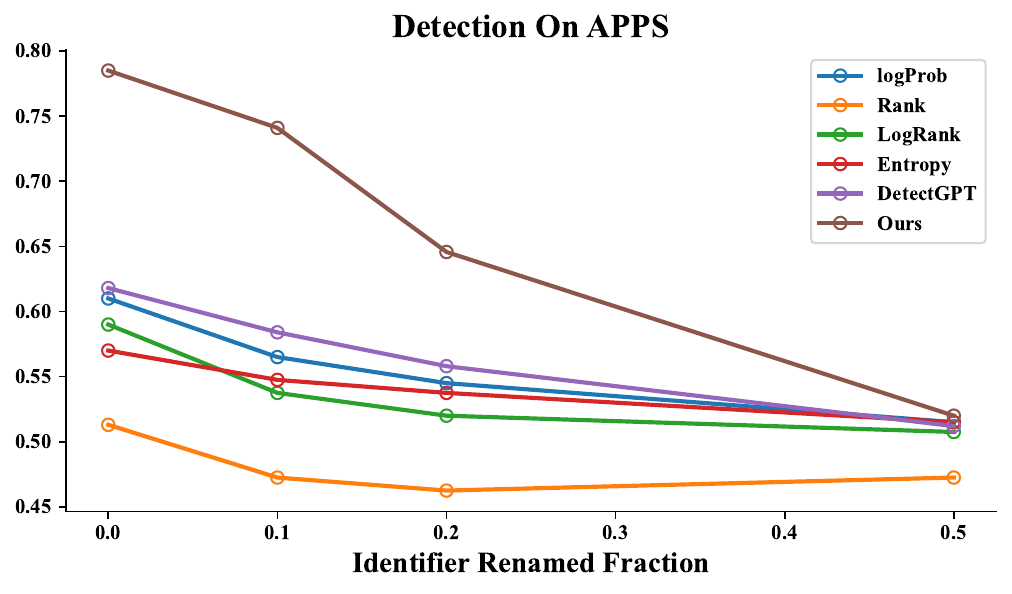}
    \caption{Detecting Revised Synthetic Code.}
    \label{fig:vary_ident_replace}
\end{figure}

\paragraph{Detecting Revised Synthetic Code.}
In real-world scenarios, developers may revise synthetic code before using it, raising the question of whether zero-shot synthetic code detectors remain effective after revisions. We focus on the most straightforward code modification: identifier renaming, which doesn't alter the code's functionality and requires no understanding of its logic. To simulate this, we extracted all identifiers and randomly replaced 10\%, 20\%, and 50\% of them with ``$var_i$". We then evaluated the zero-shot detectors on the revised synthetic code to observe performance changes as the fraction of replaced identifiers increased. The results, shown in Figure \ref{fig:vary_ident_replace}, reveal that detector performance declines as the replacement fraction increases, degrading to random guessing at 50\% replacement. However, our method consistently outperforms all others across all levels of replacement. This can be attributed to GPT-3.5-Turbo and StarChat's tendency to restore ``$var\_i$" to variable names that contain code semantics, resulting in a lower similarity between the rewritten code and the revised synthetic code compared to the similarity between the rewritten code and the original synthetic code.
One possible straightforward solution is initially using Code LLMs to restore all variable names and then applying our algorithm to detect the recovered code. We consider detection in this or more complex adversarial scenarios as a focus for future work.

\begin{figure}[!ht]
    \centering
    \includegraphics[width=0.48\textwidth]{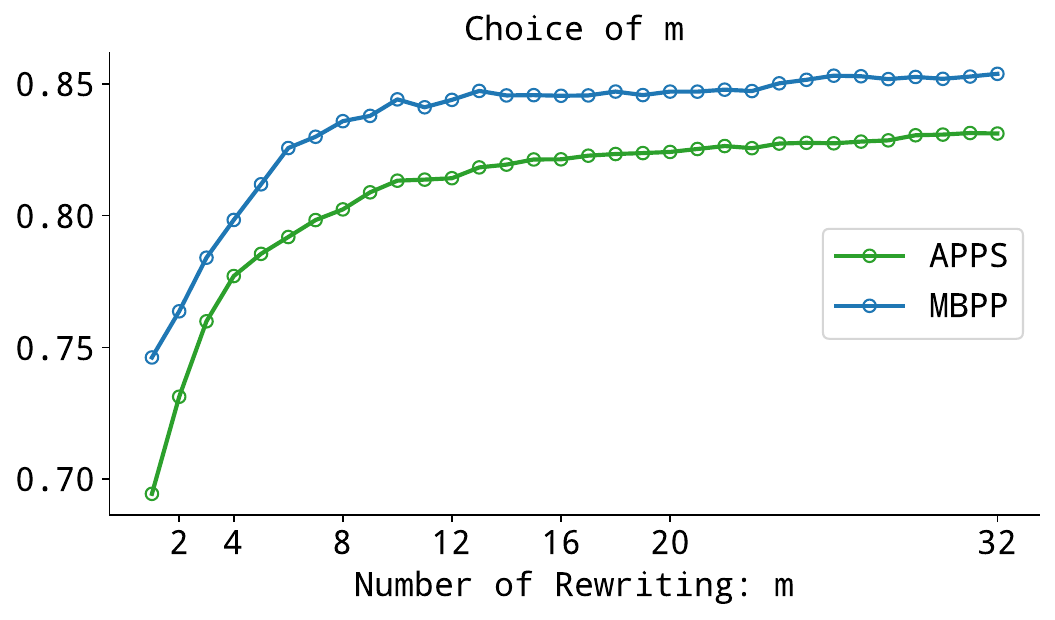}
    \caption{Choice of $m$. CodeLlama is used as Detector LLM.}
    \label{fig:impact_of_m}
\end{figure}

\paragraph{Choice of $m$.}
In our primary experiment, we set the number of rewriting $m$ in the range of $[2,4,8]$ due to limited computational resources. However, increasing $m$ can reduce randomness in code sampling and enhance the accuracy of expectation estimation. To investigate this, we conducted experiments by setting the maximum value of $m$ to 32 on the APPS and MBPP benchmarks and plotted the AUROC against changes of $m$ in Figure \ref{fig:impact_of_m}. The result indicates that the detection performance increases with $m$ with slight fluctuation and saturates around 20 - 32 rewrites.

\section{Conclusion}
In this paper, we identified the limitations of applying synthetic content detectors designed for general text to code domain and proposed a novel zero-shot synthetic code detector based on code rewriting and similarity measurement. Our approach leverages the similarity between rewritten and original code as a key indicator for detecting synthetic code. This method effectively addresses the challenges posed by the prevalence of ``low-entropy" tokens in the code domain. Extensive experiments and analyses demonstrate the superior performance and robustness of our method compared to other detectors in identifying synthetic code.

\section{Acknowledgments}
This work is supported by the Key R\&D Program of Ningbo under Grant No.2024Z115 and the Fundamental Research Funds for the Central Universities (Zhejiang University NGICS Platform).

\bibliography{aaai25}

\begin{thebibliography}{49}
\providecommand{\natexlab}[1]{#1}

\bibitem[{Achiam et~al.(2024)Achiam, Adler, Agarwal, Ahmad, Akkaya, Aleman, Almeida, Altenschmidt, Altman, Anadkat et~al.}]{achiam2023gpt}
Achiam, J.; Adler, S.; Agarwal, S.; Ahmad, L.; Akkaya, I.; Aleman, F.~L.; Almeida, D.; Altenschmidt, J.; Altman, S.; Anadkat, S.; et~al. 2024.
\newblock GPT-4 technical report.
\newblock arXiv:2303.08774.

\bibitem[{Austin et~al.(2021)Austin, Odena, Nye, Bosma, Michalewski, Dohan, Jiang, Cai, Terry, Le, and Sutton}]{austin2021mbpp}
Austin, J.; Odena, A.; Nye, M.; Bosma, M.; Michalewski, H.; Dohan, D.; Jiang, E.; Cai, C.; Terry, M.; Le, Q.; and Sutton, C. 2021.
\newblock Program Synthesis with Large Language Models.
\newblock arXiv:2108.07732.

\bibitem[{Bakhtin et~al.(2019)Bakhtin, Gross, Ott, Deng, Ranzato, and Szlam}]{bakhtin2019real}
Bakhtin, A.; Gross, S.; Ott, M.; Deng, Y.; Ranzato, M.; and Szlam, A. 2019.
\newblock Real or Fake? Learning to Discriminate Machine from Human Generated Text.
\newblock arXiv:1906.03351.

\bibitem[{Bubeck et~al.(2023)Bubeck, Chandrasekaran, Eldan, Gehrke, Horvitz, Kamar, Lee, Lee, Li, Lundberg, Nori, Palangi, Ribeiro, and Zhang}]{bubeck2023sparks}
Bubeck, S.; Chandrasekaran, V.; Eldan, R.; Gehrke, J.; Horvitz, E.; Kamar, E.; Lee, P.; Lee, Y.~T.; Li, Y.; Lundberg, S.; Nori, H.; Palangi, H.; Ribeiro, M.~T.; and Zhang, Y. 2023.
\newblock Sparks of Artificial General Intelligence: Early experiments with GPT-4.
\newblock arXiv:2303.12712.

\bibitem[{Chen et~al.(2021)Chen, Tworek, Jun, Yuan, Pinto, Kaplan, Edwards, Burda, Joseph, Brockman et~al.}]{chen2021evaluating}
Chen, M.; Tworek, J.; Jun, H.; Yuan, Q.; Pinto, H. P. d.~O.; Kaplan, J.; Edwards, H.; Burda, Y.; Joseph, N.; Brockman, G.; et~al. 2021.
\newblock Evaluating large language models trained on code.
\newblock \emph{arXiv preprint arXiv:2107.03374}.

\bibitem[{Denny et~al.(2024)Denny, Prather, Becker, Finnie-Ansley, Hellas, Leinonen, Luxton-Reilly, Reeves, Santos, and Sarsa}]{denny2024computing}
Denny, P.; Prather, J.; Becker, B.~A.; Finnie-Ansley, J.; Hellas, A.; Leinonen, J.; Luxton-Reilly, A.; Reeves, B.~N.; Santos, E.~A.; and Sarsa, S. 2024.
\newblock Computing education in the era of generative AI.
\newblock \emph{Communications of the ACM}, 67(2): 56--67.

\bibitem[{Feng et~al.(2020)Feng, Guo, Tang, Duan, Feng, Gong, Shou, Qin, Liu, Jiang, and Zhou}]{feng-etal-2020-codebert}
Feng, Z.; Guo, D.; Tang, D.; Duan, N.; Feng, X.; Gong, M.; Shou, L.; Qin, B.; Liu, T.; Jiang, D.; and Zhou, M. 2020.
\newblock {C}ode{BERT}: A Pre-Trained Model for Programming and Natural Languages.
\newblock In Cohn, T.; He, Y.; and Liu, Y., eds., \emph{Findings of the Association for Computational Linguistics: EMNLP 2020}, 1536--1547. Online: Association for Computational Linguistics.

\bibitem[{Fried et~al.(2022)Fried, Aghajanyan, Lin, Wang, Wallace, Shi, Zhong, Yih, Zettlemoyer, and Lewis}]{fried2022incoder}
Fried, D.; Aghajanyan, A.; Lin, J.; Wang, S.; Wallace, E.; Shi, F.; Zhong, R.; Yih, W.-t.; Zettlemoyer, L.; and Lewis, M. 2022.
\newblock Incoder: A generative model for code infilling and synthesis.
\newblock \emph{arXiv preprint arXiv:2204.05999}.

\bibitem[{Gao, Yao, and Chen(2021)}]{gao2021simcse}
Gao, T.; Yao, X.; and Chen, D. 2021.
\newblock {S}im{CSE}: Simple Contrastive Learning of Sentence Embeddings.
\newblock In \emph{Proceedings of the 2021 Conference on Empirical Methods in Natural Language Processing}, 6894--6910. Association for Computational Linguistics.

\bibitem[{Gehrmann, Strobelt, and Rush(2019)}]{gehrmann-etal-2019-gltr}
Gehrmann, S.; Strobelt, H.; and Rush, A. 2019.
\newblock {GLTR}: Statistical Detection and Visualization of Generated Text.
\newblock In Costa-juss{\`a}, M.~R.; and Alfonseca, E., eds., \emph{Proceedings of the 57th Annual Meeting of the Association for Computational Linguistics: System Demonstrations}, 111--116. Florence, Italy: Association for Computational Linguistics.

\bibitem[{{Github}(2021)}]{copilot}
{Github}. 2021.
\newblock Github Copilot.
\newblock \url{https://copilot.github.com}.
\newblock Accessed: 2018-12-06.

\bibitem[{{GPTZero}(2023)}]{gptzero}
{GPTZero}. 2023.
\newblock GPTZero.
\newblock \url{https://gptzero.me/}.
\newblock Accessed: 2023-12-06.

\bibitem[{Greene et~al.(2022)Greene, Sanders, Weng, and Neelakantan}]{openaiemb2022}
Greene, R.; Sanders; Weng, L.; and Neelakantan, A. 2022.
\newblock New and improved embedding model.

\bibitem[{Guo et~al.(2022)Guo, Lu, Duan, Wang, Zhou, and Yin}]{guo-etal-2022-unixcoder}
Guo, D.; Lu, S.; Duan, N.; Wang, Y.; Zhou, M.; and Yin, J. 2022.
\newblock {U}ni{X}coder: Unified Cross-Modal Pre-training for Code Representation.
\newblock In Muresan, S.; Nakov, P.; and Villavicencio, A., eds., \emph{Proceedings of the 60th Annual Meeting of the Association for Computational Linguistics (Volume 1: Long Papers)}, 7212--7225. Dublin, Ireland: Association for Computational Linguistics.

\bibitem[{Guo et~al.(2021)Guo, Ren, Lu, Feng, Tang, LIU, Zhou, Duan, Svyatkovskiy, Fu, Tufano, Deng, Clement, Drain, Sundaresan, Yin, Jiang, and Zhou}]{guo2021graphcodebert}
Guo, D.; Ren, S.; Lu, S.; Feng, Z.; Tang, D.; LIU, S.; Zhou, L.; Duan, N.; Svyatkovskiy, A.; Fu, S.; Tufano, M.; Deng, S.~K.; Clement, C.; Drain, D.; Sundaresan, N.; Yin, J.; Jiang, D.; and Zhou, M. 2021.
\newblock GraphCode{\{}BERT{\}}: Pre-training Code Representations with Data Flow.
\newblock In \emph{International Conference on Learning Representations}.

\bibitem[{Guo et~al.(2024)Guo, Zhu, Dejian~Yang, Kai~Dong, Chen, Bi, Wu, Li, Luo, Xiong, and Liang}]{deepseek-coder}
Guo, D.; Zhu, Q.; Dejian~Yang, Z.~X.; Kai~Dong, W.~Z.; Chen, G.; Bi, X.; Wu, Y.; Li, Y.; Luo, F.; Xiong, Y.; and Liang, W. 2024.
\newblock DeepSeek-Coder: When the Large Language Model Meets Programming -- The Rise of Code Intelligence.

\bibitem[{He and Vechev(2023)}]{he2023large}
He, J.; and Vechev, M. 2023.
\newblock Large language models for code: Security hardening and adversarial testing.
\newblock In \emph{Proceedings of the 2023 ACM SIGSAC Conference on Computer and Communications Security}, 1865--1879.

\bibitem[{Hendrycks et~al.(2021)Hendrycks, Basart, Kadavath, Mazeika, Arora, Guo, Burns, Puranik, He, Song, and Steinhardt}]{hendrycks2021apps}
Hendrycks, D.; Basart, S.; Kadavath, S.; Mazeika, M.; Arora, A.; Guo, E.; Burns, C.; Puranik, S.; He, H.; Song, D.; and Steinhardt, J. 2021.
\newblock Measuring Coding Challenge Competence With {APPS}.
\newblock In \emph{Thirty-fifth Conference on Neural Information Processing Systems Datasets and Benchmarks Track (Round 2)}.

\bibitem[{Husain et~al.(2020)Husain, Wu, Gazit, Allamanis, and Brockschmidt}]{husain2020codesearchnet}
Husain, H.; Wu, H.-H.; Gazit, T.; Allamanis, M.; and Brockschmidt, M. 2020.
\newblock CodeSearchNet Challenge: Evaluating the State of Semantic Code Search.
\newblock arXiv:1909.09436.

\bibitem[{Ippolito et~al.(2020)Ippolito, Duckworth, Callison-Burch, and Eck}]{ippolito-etal-2020-automatic}
Ippolito, D.; Duckworth, D.; Callison-Burch, C.; and Eck, D. 2020.
\newblock Automatic Detection of Generated Text is Easiest when Humans are Fooled.
\newblock In Jurafsky, D.; Chai, J.; Schluter, N.; and Tetreault, J., eds., \emph{Proceedings of the 58th Annual Meeting of the Association for Computational Linguistics}, 1808--1822. Online: Association for Computational Linguistics.

\bibitem[{Kazemitabaar et~al.(2023)Kazemitabaar, Hou, Henley, Ericson, Weintrop, and Grossman}]{kazemitabaar2023novices}
Kazemitabaar, M.; Hou, X.; Henley, A.; Ericson, B.~J.; Weintrop, D.; and Grossman, T. 2023.
\newblock How novices use LLM-based code generators to solve CS1 coding tasks in a self-paced learning environment.
\newblock In \emph{Proceedings of the 23rd Koli Calling International Conference on Computing Education Research}, 1--12.

\bibitem[{Li et~al.(2023)Li, allal, Zi, Muennighoff, Kocetkov, Mou, Marone, Akiki, LI, Chim, Liu, Zheltonozhskii, Zhuo, Wang, Dehaene, Lamy-Poirier, Monteiro, Gontier, Yee, Umapathi, Zhu, Lipkin, Oblokulov, Wang, Murthy, Stillerman, Patel, Abulkhanov, Zocca, Dey, Zhang, Bhattacharyya, Yu, Luccioni, Villegas, Zhdanov, Lee, Timor, Ding, Schlesinger, Schoelkopf, Ebert, Dao, Mishra, Gu, Anderson, Dolan-Gavitt, Contractor, Reddy, Fried, Bahdanau, Jernite, Ferrandis, Hughes, Wolf, Guha, Werra, and de~Vries}]{li2023starcoder}
Li, R.; allal, L.~B.; Zi, Y.; Muennighoff, N.; Kocetkov, D.; Mou, C.; Marone, M.; Akiki, C.; LI, J.; Chim, J.; Liu, Q.; Zheltonozhskii, E.; Zhuo, T.~Y.; Wang, T.; Dehaene, O.; Lamy-Poirier, J.; Monteiro, J.; Gontier, N.; Yee, M.-H.; Umapathi, L.~K.; Zhu, J.; Lipkin, B.; Oblokulov, M.; Wang, Z.; Murthy, R.; Stillerman, J.~T.; Patel, S.~S.; Abulkhanov, D.; Zocca, M.; Dey, M.; Zhang, Z.; Bhattacharyya, U.; Yu, W.; Luccioni, S.; Villegas, P.; Zhdanov, F.; Lee, T.; Timor, N.; Ding, J.; Schlesinger, C.~S.; Schoelkopf, H.; Ebert, J.; Dao, T.; Mishra, M.; Gu, A.; Anderson, C.~J.; Dolan-Gavitt, B.; Contractor, D.; Reddy, S.; Fried, D.; Bahdanau, D.; Jernite, Y.; Ferrandis, C.~M.; Hughes, S.; Wolf, T.; Guha, A.; Werra, L.~V.; and de~Vries, H. 2023.
\newblock StarCoder: may the source be with you!
\newblock \emph{Transactions on Machine Learning Research}.
\newblock Reproducibility Certification.

\bibitem[{Li et~al.(2022{\natexlab{a}})Li, Guo, Gong, Lin, Shen, Qiu, Jiang, Chen, and Duan}]{li-etal-2022-soft}
Li, X.; Guo, D.; Gong, Y.; Lin, Y.; Shen, Y.; Qiu, X.; Jiang, D.; Chen, W.; and Duan, N. 2022{\natexlab{a}}.
\newblock Soft-Labeled Contrastive Pre-Training for Function-Level Code Representation.
\newblock In Goldberg, Y.; Kozareva, Z.; and Zhang, Y., eds., \emph{Findings of the Association for Computational Linguistics: EMNLP 2022}, 118--129. Abu Dhabi, United Arab Emirates: Association for Computational Linguistics.

\bibitem[{Li et~al.(2022{\natexlab{b}})Li, Choi, Chung, Kushman, Schrittwieser, Leblond, Eccles, Keeling, Gimeno, Dal~Lago, Hubert, Choy, de~Masson~d’Autume, Babuschkin, Chen, Huang, Welbl, Gowal, Cherepanov, Molloy, Mankowitz, Sutherland~Robson, Kohli, de~Freitas, Kavukcuoglu, and Vinyals}]{alphacode_li_2022}
Li, Y.; Choi, D.; Chung, J.; Kushman, N.; Schrittwieser, J.; Leblond, R.; Eccles, T.; Keeling, J.; Gimeno, F.; Dal~Lago, A.; Hubert, T.; Choy, P.; de~Masson~d’Autume, C.; Babuschkin, I.; Chen, X.; Huang, P.-S.; Welbl, J.; Gowal, S.; Cherepanov, A.; Molloy, J.; Mankowitz, D.~J.; Sutherland~Robson, E.; Kohli, P.; de~Freitas, N.; Kavukcuoglu, K.; and Vinyals, O. 2022{\natexlab{b}}.
\newblock Competition-level code generation with AlphaCode.
\newblock \emph{Science}, 378(6624): 1092–1097.

\bibitem[{Liu et~al.(2019)Liu, Ott, Goyal, Du, Joshi, Chen, Levy, Lewis, Zettlemoyer, and Stoyanov}]{liu2019roberta}
Liu, Y.; Ott, M.; Goyal, N.; Du, J.; Joshi, M.; Chen, D.; Levy, O.; Lewis, M.; Zettlemoyer, L.; and Stoyanov, V. 2019.
\newblock RoBERTa: A Robustly Optimized BERT Pretraining Approach.
\newblock arXiv:1907.11692.

\bibitem[{Mao et~al.(2024)Mao, Vondrick, Wang, and Yang}]{mao2024raidar}
Mao, C.; Vondrick, C.; Wang, H.; and Yang, J. 2024.
\newblock Raidar: geneRative {AI} Detection viA Rewriting.
\newblock In \emph{The Twelfth International Conference on Learning Representations}.

\bibitem[{Mitchell et~al.(2023)Mitchell, Lee, Khazatsky, Manning, and Finn}]{mitchell2023detectgpt}
Mitchell, E.; Lee, Y.; Khazatsky, A.; Manning, C.~D.; and Finn, C. 2023.
\newblock {D}etect{GPT}: Zero-Shot Machine-Generated Text Detection using Probability Curvature.
\newblock In Krause, A.; Brunskill, E.; Cho, K.; Engelhardt, B.; Sabato, S.; and Scarlett, J., eds., \emph{Proceedings of the 40th International Conference on Machine Learning}, volume 202 of \emph{Proceedings of Machine Learning Research}, 24950--24962. PMLR.

\bibitem[{Nijkamp et~al.(2022)Nijkamp, Pang, Hayashi, Tu, Wang, Zhou, Savarese, and Xiong}]{nijkamp2022codegen}
Nijkamp, E.; Pang, B.; Hayashi, H.; Tu, L.; Wang, H.; Zhou, Y.; Savarese, S.; and Xiong, C. 2022.
\newblock Codegen: An open large language model for code with multi-turn program synthesis.
\newblock \emph{arXiv preprint arXiv:2203.13474}.

\bibitem[{OpenAI(2019)}]{openai_detector}
OpenAI. 2019.
\newblock gpt-2-output-dataset.

\bibitem[{OpenAI(2022)}]{openai2022gpt35}
OpenAI. 2022.
\newblock Introducing ChatGPT.

\bibitem[{Pearce et~al.(2022)Pearce, Ahmad, Tan, Dolan-Gavitt, and Karri}]{pearce2022asleep}
Pearce, H.; Ahmad, B.; Tan, B.; Dolan-Gavitt, B.; and Karri, R. 2022.
\newblock Asleep at the keyboard? assessing the security of github copilot’s code contributions.
\newblock In \emph{2022 IEEE Symposium on Security and Privacy (SP)}, 754--768. IEEE.

\bibitem[{Pu et~al.(2023)Pu, Sarwar, Abdullah, Rehman, Kim, Bhattacharya, Javed, and Viswanath}]{pu2023deepfake}
Pu, J.; Sarwar, Z.; Abdullah, S.~M.; Rehman, A.; Kim, Y.; Bhattacharya, P.; Javed, M.; and Viswanath, B. 2023.
\newblock Deepfake text detection: Limitations and opportunities.
\newblock In \emph{2023 IEEE Symposium on Security and Privacy (SP)}, 1613--1630. IEEE.

\bibitem[{Raffel et~al.(2020)Raffel, Shazeer, Roberts, Lee, Narang, Matena, Zhou, Li, and Liu}]{raffel-t5}
Raffel, C.; Shazeer, N.; Roberts, A.; Lee, K.; Narang, S.; Matena, M.; Zhou, Y.; Li, W.; and Liu, P.~J. 2020.
\newblock Exploring the Limits of Transfer Learning with a Unified Text-to-Text Transformer.
\newblock \emph{J. Mach. Learn. Res.}, 21(1).

\bibitem[{Rozière et~al.(2023)Rozière, Gehring, Gloeckle, Sootla, Gat, Tan, Adi, Liu, Remez, Rapin, Kozhevnikov, Evtimov, Bitton, Bhatt, Ferrer, Grattafiori, Xiong, Défossez, Copet, Azhar, Touvron, Martin, Usunier, Scialom, and Synnaeve}]{codellama2023}
Rozière, B.; Gehring, J.; Gloeckle, F.; Sootla, S.; Gat, I.; Tan, X.~E.; Adi, Y.; Liu, J.; Remez, T.; Rapin, J.; Kozhevnikov, A.; Evtimov, I.; Bitton, J.; Bhatt, M.; Ferrer, C.~C.; Grattafiori, A.; Xiong, W.; Défossez, A.; Copet, J.; Azhar, F.; Touvron, H.; Martin, L.; Usunier, N.; Scialom, T.; and Synnaeve, G. 2023.
\newblock Code Llama: Open Foundation Models for Code.
\newblock arXiv:2308.12950.

\bibitem[{StackOverflow(2023{\natexlab{a}})}]{stackoverflow-survey}
StackOverflow. 2023{\natexlab{a}}.
\newblock Developer sentiment around AI/ML.

\bibitem[{StackOverflow(2023{\natexlab{b}})}]{stackoverflow-developer}
StackOverflow. 2023{\natexlab{b}}.
\newblock Stack Overflow Developer Survey 2023.

\bibitem[{Su et~al.(2023)Su, Zhuo, Wang, and Nakov}]{su2023detectllm}
Su, J.; Zhuo, T.~Y.; Wang, D.; and Nakov, P. 2023.
\newblock Detectllm: Leveraging log rank information for zero-shot detection of machine-generated text.
\newblock \emph{arXiv preprint arXiv:2306.05540}.

\bibitem[{Touvron et~al.(2023{\natexlab{a}})Touvron, Lavril, Izacard, Martinet, Lachaux, Lacroix, Rozière, Goyal, Hambro, Azhar, Rodriguez, Joulin, Grave, and Lample}]{touvron2023llamaopenefficientfoundation}
Touvron, H.; Lavril, T.; Izacard, G.; Martinet, X.; Lachaux, M.-A.; Lacroix, T.; Rozière, B.; Goyal, N.; Hambro, E.; Azhar, F.; Rodriguez, A.; Joulin, A.; Grave, E.; and Lample, G. 2023{\natexlab{a}}.
\newblock LLaMA: Open and Efficient Foundation Language Models.
\newblock arXiv:2302.13971.

\bibitem[{Touvron et~al.(2023{\natexlab{b}})Touvron, Martin, Stone, Albert, Almahairi, Babaei, Bashlykov, Batra, Bhargava, Bhosale, Bikel, Blecher, Ferrer, Chen, Cucurull, Esiobu, Fernandes, Fu, Fu, Fuller, Gao, Goswami, Goyal, Hartshorn, Hosseini, Hou, Inan, Kardas, Kerkez, Khabsa, Kloumann, Korenev, Koura, Lachaux, Lavril, Lee, Liskovich, Lu, Mao, Martinet, Mihaylov, Mishra, Molybog, Nie, Poulton, Reizenstein, Rungta, Saladi, Schelten, Silva, Smith, Subramanian, Tan, Tang, Taylor, Williams, Kuan, Xu, Yan, Zarov, Zhang, Fan, Kambadur, Narang, Rodriguez, Stojnic, Edunov, and Scialom}]{touvron2023llama}
Touvron, H.; Martin, L.; Stone, K.; Albert, P.; Almahairi, A.; Babaei, Y.; Bashlykov, N.; Batra, S.; Bhargava, P.; Bhosale, S.; Bikel, D.; Blecher, L.; Ferrer, C.~C.; Chen, M.; Cucurull, G.; Esiobu, D.; Fernandes, J.; Fu, J.; Fu, W.; Fuller, B.; Gao, C.; Goswami, V.; Goyal, N.; Hartshorn, A.; Hosseini, S.; Hou, R.; Inan, H.; Kardas, M.; Kerkez, V.; Khabsa, M.; Kloumann, I.; Korenev, A.; Koura, P.~S.; Lachaux, M.-A.; Lavril, T.; Lee, J.; Liskovich, D.; Lu, Y.; Mao, Y.; Martinet, X.; Mihaylov, T.; Mishra, P.; Molybog, I.; Nie, Y.; Poulton, A.; Reizenstein, J.; Rungta, R.; Saladi, K.; Schelten, A.; Silva, R.; Smith, E.~M.; Subramanian, R.; Tan, X.~E.; Tang, B.; Taylor, R.; Williams, A.; Kuan, J.~X.; Xu, P.; Yan, Z.; Zarov, I.; Zhang, Y.; Fan, A.; Kambadur, M.; Narang, S.; Rodriguez, A.; Stojnic, R.; Edunov, S.; and Scialom, T. 2023{\natexlab{b}}.
\newblock Llama 2: Open Foundation and Fine-Tuned Chat Models.
\newblock arXiv:2307.09288.

\bibitem[{Tunstall et~al.(2023)Tunstall, Lambert, Rajani, Beeching, Le~Scao, von Werra, Han, Schmid, and Rush}]{Tunstall2023starchat-alpha}
Tunstall, L.; Lambert, N.; Rajani, N.; Beeching, E.; Le~Scao, T.; von Werra, L.; Han, S.; Schmid, P.; and Rush, A. 2023.
\newblock Creating a Coding Assistant with StarCoder.
\newblock \emph{Hugging Face Blog}.
\newblock Https://huggingface.co/blog/starchat.

\bibitem[{Uchendu et~al.(2020)Uchendu, Le, Shu, and Lee}]{uchendu2020authorship}
Uchendu, A.; Le, T.; Shu, K.; and Lee, D. 2020.
\newblock Authorship attribution for neural text generation.
\newblock In \emph{Proceedings of the 2020 Conference on Empirical Methods in Natural Language Processing (EMNLP)}, 8384--8395.

\bibitem[{Wang et~al.(2021{\natexlab{a}})Wang, Wang, Mi, Zhou, Wan, Liu, Li, Wu, Liu, and Jiang}]{wang2021syncobert}
Wang, X.; Wang, Y.; Mi, F.; Zhou, P.; Wan, Y.; Liu, X.; Li, L.; Wu, H.; Liu, J.; and Jiang, X. 2021{\natexlab{a}}.
\newblock SynCoBERT: Syntax-Guided Multi-Modal Contrastive Pre-Training for Code Representation.
\newblock arXiv:2108.04556.

\bibitem[{Wang et~al.(2022)Wang, Wang, Wan, Wang, Zhou, Li, Wu, and Liu}]{wang-etal-2022-code}
Wang, X.; Wang, Y.; Wan, Y.; Wang, J.; Zhou, P.; Li, L.; Wu, H.; and Liu, J. 2022.
\newblock {CODE}-{MVP}: Learning to Represent Source Code from Multiple Views with Contrastive Pre-Training.
\newblock In Carpuat, M.; de~Marneffe, M.-C.; and Meza~Ruiz, I.~V., eds., \emph{Findings of the Association for Computational Linguistics: NAACL 2022}, 1066--1077. Seattle, United States: Association for Computational Linguistics.

\bibitem[{Wang et~al.(2021{\natexlab{b}})Wang, Wang, Joty, and Hoi}]{wang2021codet5}
Wang, Y.; Wang, W.; Joty, S.; and Hoi, S.~C. 2021{\natexlab{b}}.
\newblock {C}ode{T}5: Identifier-aware Unified Pre-trained Encoder-Decoder Models for Code Understanding and Generation.
\newblock In Moens, M.-F.; Huang, X.; Specia, L.; and Yih, S. W.-t., eds., \emph{Proceedings of the 2021 Conference on Empirical Methods in Natural Language Processing}, 8696--8708. Online and Punta Cana, Dominican Republic: Association for Computational Linguistics.

\bibitem[{Wei et~al.(2022)Wei, Wang, Schuurmans, Bosma, ichter, Xia, Chi, Le, and Zhou}]{wei2022chainofthought}
Wei, J.; Wang, X.; Schuurmans, D.; Bosma, M.; ichter, b.; Xia, F.; Chi, E.; Le, Q.~V.; and Zhou, D. 2022.
\newblock Chain-of-Thought Prompting Elicits Reasoning in Large Language Models.
\newblock In Koyejo, S.; Mohamed, S.; Agarwal, A.; Belgrave, D.; Cho, K.; and Oh, A., eds., \emph{Advances in Neural Information Processing Systems}, volume~35, 24824--24837. Curran Associates, Inc.

\bibitem[{Ye et~al.(2021)Ye, Zhou, Venkat, Marcus, Tatbul, Tithi, Hasabnis, Petersen, Mattson, Kraska, Dubey, Sarkar, and Gottschlich}]{ye2021misim}
Ye, F.; Zhou, S.; Venkat, A.; Marcus, R.; Tatbul, N.; Tithi, J.~J.; Hasabnis, N.; Petersen, P.; Mattson, T.; Kraska, T.; Dubey, P.; Sarkar, V.; and Gottschlich, J. 2021.
\newblock MISIM: A Neural Code Semantics Similarity System Using the Context-Aware Semantics Structure.
\newblock arXiv:2006.05265.

\bibitem[{Zellers et~al.(2019)Zellers, Holtzman, Rashkin, Bisk, Farhadi, Roesner, and Choi}]{zellers2019defending}
Zellers, R.; Holtzman, A.; Rashkin, H.; Bisk, Y.; Farhadi, A.; Roesner, F.; and Choi, Y. 2019.
\newblock Defending against neural fake news.
\newblock \emph{Advances in neural information processing systems}, 32.

\bibitem[{Zheng et~al.(2023)Zheng, Xia, Zou, Dong, Wang, Xue, Wang, Shen, Wang, Li et~al.}]{zheng2023codegeex}
Zheng, Q.; Xia, X.; Zou, X.; Dong, Y.; Wang, S.; Xue, Y.; Wang, Z.; Shen, L.; Wang, A.; Li, Y.; et~al. 2023.
\newblock Codegeex: A pre-trained model for code generation with multilingual evaluations on humaneval-x.
\newblock \emph{arXiv preprint arXiv:2303.17568}.

\bibitem[{Zhong et~al.(2020)Zhong, Tang, Xu, Wang, Duan, Zhou, Wang, and Yin}]{zhong-etal-2020-neural}
Zhong, W.; Tang, D.; Xu, Z.; Wang, R.; Duan, N.; Zhou, M.; Wang, J.; and Yin, J. 2020.
\newblock Neural Deepfake Detection with Factual Structure of Text.
\newblock In Webber, B.; Cohn, T.; He, Y.; and Liu, Y., eds., \emph{Proceedings of the 2020 Conference on Empirical Methods in Natural Language Processing (EMNLP)}, 2461--2470. Online: Association for Computational Linguistics.

\end{thebibliography}

\end{document}